\documentclass[aps,prl,twocolumn,showpacs]{revtex4}%
\usepackage{amsfonts}
\usepackage{amsmath}
\usepackage{amssymb}
\usepackage{graphicx}%
\providecommand{\U}[1]{\protect\rule{.1in}{.1in}}
\begin{document}
\title[Thermalization of a nonequilibrium electron-positron-photon plasma]{Thermalization of a nonequilibrium electron-positron-photon plasma}
\author{A.G. Aksenov}
\altaffiliation{Institute for Theoretical and Experimental Physics, B. Cheremushkinskaya 25,
117218 Moscow, Russia}

\author{R. Ruffini}
\author{G.V. Vereshchagin}
\affiliation{ICRANet p.le della Repubblica, 10, 65100 Pescara,\ Italy}
\affiliation{ICRA and University of Rome \textquotedblleft Sapienza\textquotedblright,
Physics Department, p.le A. Moro 5, 00185 Rome, Italy}
\keywords{Electron-positron plasmas; Kinetic theory}
\pacs{52.27.Ep; 05.20.Dd}

\begin{abstract}
Starting from a nonequilibrium configuration we analyse the essential role of
the direct and the inverse binary and triple interactions in reaching an
asymptotic thermal equilibrium in a homogeneous isotropic
electron-positron-photon plasma. We focus on energies in the range 0.1--10
MeV. We numerically integrate the integro-partial differential relativistic
Boltzmann equation with the exact QED collisional integrals taking into
account all binary and triple interactions in the plasma. We show that first,
when detailed balance is reached for all binary interactions on a timescale
$t_{k}\lesssim10^{-14}$sec, photons and electron-positron pairs establish
kinetic equilibrium. Successively, when triple interactions fulfill the
detailed balance on a timescale $t_{eq}\lesssim10^{-12}$sec, the plasma
reaches thermal equilibrium. It is shown that neglecting the inverse triple
interactions prevents reaching thermal equilibrium. Our results obtained in
the theoretical physics domain also find application in astrophysics and cosmology.

\end{abstract}
\maketitle

An electron-positron plasma is of interest in many fields of physics and
astrophysics: the early universe \cite{Kolb1990}, gamma-ray bursts
\cite{Piran1999}, active galactic nuclei \cite{Wardle1998}, the center of our
Galaxy \cite{Churazov2005},\ hypothetical quark stars \cite{Usov1998} and
ultraintense lasers \cite{Blaschke2006}.

A detailed study of the relevant processes and possible equilibrium
configurations in an optically thin pair plasma are given in
\cite{Bisnovatyi1971}. In all the above-mentioned applications the precise
knowledge of the optically thick plasma evolution is required. In this case
there exists only a qualitative description and an assumption of thermal
equilibrium is often adopted without explicit proof \cite{Piran1999}.

In this Letter we consider a uniform isotropic electron-positron-photon plasma
in the absence of external electromagnetic fields and we describe its
evolution starting from arbitrary nonequilibrium initial conditions up to
reaching thermal equilibrium. We are interested in the range of final
temperatures in thermal equilibrium, bracketing the electron rest mass energy%
\begin{equation}
0.1\;\mathrm{MeV}\lesssim T_{th}\lesssim10\;\mathrm{MeV.} \label{temp}%
\end{equation}
These boundaries are required for the study of electron-positron pairs in
absence of the production of other particles such as muons. We assume that the
energy density of the plasma is constant and is, correspondingly, in the range
$1.6\;10^{22}\frac{\mathrm{erg}}{\mathrm{cm}^{3}}<\rho<3.8\;10^{30}%
\frac{\mathrm{erg}}{\mathrm{cm}^{3}}$. The relative number densities at
thermal equilibrium will be $3.1\;10^{28}\mathrm{cm}^{-3}<n_{th}%
<7.9\;10^{34}\mathrm{cm}^{-3}$.

We adopt a kinetic description for the distribution functions of electrons,
positrons and photons. In our case the plasma parameter is small,
$g=(nr_{D}^{3})^{-1}\ll1$, where $r_{D}$ is the Debye length, and therefore we
use one-particle distribution functions. Besides, in our case electrons and
positrons are non-degenerate. We solve numerically the relativistic Boltzmann
equations \cite{BB1956} which for homogeneous and isotropic distribution
functions of electrons, positrons and photons reduce to%
\begin{equation}
\frac{1}{c}\frac{\partial f_{i}}{\partial t}=\sum_{q}\left(  \eta_{i}^{q}%
-\chi_{i}^{q}f_{i}\right)  ,\label{BE}%
\end{equation}
where $f_{i}(\epsilon,t)$ are their distribution functions, the index $i$
denotes the type of particle, $\epsilon$ is their energy, and $\eta_{i}^{q}$
and $\chi_{i}^{q}$ are the emission and the absorption coefficients for the
production of a particle of type \textquotedblleft$i$" via the physical
process labeled by $q$.

In order to solve equations (\ref{BE}) we use a finite difference method by
introducing a computational grid in the phase space to represent the
distribution functions and to compute collisional integrals \cite{Aksenov2004}%
. The result of this procedure is the stiff system of ordinary differential
equations to be solved with the implicit Gear method \cite{Hall1976}. For
binary interactions we use exact QED matrix elements \cite{Berestetskii1982}.
For triple interactions we compute emission and absorption coefficients
following Svensson \cite{Svensson1984}. The Compton scattering of photons, for
instance, is described by \cite{Aksenov2004}
\begin{align}
\eta_{\gamma}^{\mathrm{cs}} &  =%
{\displaystyle\int}
d\mathbf{k}^{\prime}d\mathbf{p}d\mathbf{p}^{\prime}w_{\mathbf{k}^{\prime
}\mathbf{p}^{\prime};\mathbf{k},\mathbf{p}}f_{\gamma}(\mathbf{k}^{\prime
},t)f_{\pm}(\mathbf{p}^{\prime},t),\label{chics}\\
\chi_{\gamma}^{\mathrm{cs}}f_{\gamma} &  =%
{\displaystyle\int}
d\mathbf{k}^{\prime}d\mathbf{p}d\mathbf{p}^{\prime}w_{\mathbf{k}^{\prime
}\mathbf{p}^{\prime};\mathbf{k},\mathbf{p}}f_{\gamma}(\mathbf{k},t)f_{\pm
}(\mathbf{p},t),\label{etacs}%
\end{align}
where
\begin{equation}
w_{\mathbf{k}^{\prime}\mathbf{p}^{\prime};\mathbf{k},\mathbf{p}}=\frac
{1}{\left(  2\pi\hbar\right)  ^{2}}\delta^{4}(k+p-k^{\prime}-p^{\prime}%
)\frac{\left\vert M_{fi}\right\vert ^{2}}{16\epsilon_{\gamma}\epsilon_{\pm
}\epsilon_{\gamma}^{\prime}\epsilon_{\pm}^{\prime}}\label{prob}%
\end{equation}
is the corresponding transition probability, $k=(\epsilon_{\gamma
}/c,\mathbf{k})$ and $p=(\epsilon_{e}/c,\mathbf{p})$ are four-momenta of
photon and positron (electron), primes denote particles after the interaction,
and $M_{fi}$ is the matrix element for the considered process.

For such a dense plasma collisional integrals in (\ref{BE})\ should include
not only binary interactions, having order $\alpha$ in Feynmann diagrams,
where $\alpha$ is the fine structure constant, but also triple ones, having
order $\alpha^{2}$ \cite{Berestetskii1982}. We consider all possible binary
and triple interactions between electrons, positrons and photons as summarized
in Tab.~\ref{tab1}.%

\begin{table}[tbp] \centering
\begin{tabular}
[c]{|c|c|}\hline
Binary interactions & Radiative variants\\\hline\hline
{M{\o }ller, Bhabha} & {Bremsstrahlung}\\
{$e^{\pm}{e^{\pm\prime}\leftrightarrow e^{\pm}}^{\prime\prime}$}${e^{\pm}%
}^{\prime\prime\prime}$ & {$e^{\pm}e^{\pm\prime}{\leftrightarrow}e^{\pm
\prime\prime}e^{\pm\prime\prime\prime}\gamma$}\\
{$e^{\pm}{e^{\mp}\leftrightarrow e^{\pm\prime}}$}${e^{\mp}}^{\prime}$ &
{$e^{\pm}e^{\mp}{\leftrightarrow}e^{\pm\prime}e{^{\mp\prime}}\gamma$}\\\hline
Single {Compton} & {Double Compton}\\
{ $e^{\pm}\gamma{\leftrightarrow}e^{\pm}\gamma^{\prime}$} & {$e^{\pm}%
\gamma{\leftrightarrow}e^{\pm\prime}\gamma^{\prime}\gamma^{\prime\prime}$%
}\\\hline
{Pair production} & Radiative pair production\\
and annihilation & and three photon annihilation\\
{$\gamma\gamma^{\prime}{\leftrightarrow}e^{\pm}e^{\mp}$} & $\gamma
\gamma^{\prime}${${\leftrightarrow}e^{\pm}e^{\mp}$}$\gamma^{\prime\prime}$\\
& $e^{\pm}\gamma${${\leftrightarrow}e^{\pm\prime}{e^{\mp}}e^{\pm\prime\prime}%
$}\\
& {$e^{\pm}e^{\mp}{\leftrightarrow}\gamma\gamma^{\prime}$}$\gamma
^{\prime\prime}$\\\hline
\end{tabular}
\caption{Microphysical processes in the pair plasma.}\label{tab1}%
\end{table}%
Each of the above reactions is characterized by the corresponding timescale
and optical depth. For Compton scattering, for instance, we have%
\begin{equation}
t_{\mathrm{cs}}=\frac{1}{\sigma_{T}n_{\pm}c},\qquad\tau_{\mathrm{cs}}%
=\sigma_{T}n_{\pm}R_{0},\label{ttau}%
\end{equation}
where $\sigma_{T}$ is the Thomson cross-section and $n_{\pm}$ is the number
density of pairs. There are two timescales in our problem that characterize
the condition of detailed balance between direct and inverse reactions, $\sim
t_{\mathrm{cs}}$ for binary and $\alpha^{-1}t_{\mathrm{cs}}$ for triple
interactions respectively. In the first phase of the system evolution the
binary interactions are found to have a predominant role. Starting from
arbitrary distribution functions we find a common development: at the time
$t_{\mathrm{cs}}$ the distribution functions always have evolved in a
functional form on the entire energy range, depending only on two parameters.
We find in fact for the distribution functions the expressions
\begin{equation}
f_{i}(\varepsilon)=\exp\left(  -\frac{\varepsilon-\varphi_{i}}{\theta_{i}%
}\right)  ,\label{dk}%
\end{equation}
with chemical potential $\varphi_{i}\equiv\frac{\mu_{i}}{m_{e}c^{2}}$ and
temperature $\theta_{i}\equiv\frac{k_{B}T_{i}}{m_{e}c^{2}}$, where
$\varepsilon\equiv\frac{\epsilon}{m_{e}c^{2}}$ is the energy of the particle,
$m_{e}$ is the electron mass and $k_{B}$ is Boltzmann's constant. Such a
configuration corresponds to a kinetic equilibrium \cite{Pilla1997}%
,\cite{Ehlers1973} in which electrons, positrons and photons acquire a common
temperature and nonzero chemical potential. At the same time we found that
triple interactions become essential for $t>t_{\mathrm{cs}}$, after the
establishment of kinetic equilibrium. Such triple interactions, both direct
and inverse, are indeed essential in achieving the thermal equilibrium.

In (\ref{dk}) analogously to the temperature, defining the average kinetic
energy in the system, the chemical potential represents deviation from the
thermal equilibrium through the relation $\varphi=\theta\ln(n/n_{eq})$, where
$n_{eq}$ are concentrations of particles in thermal equilibrium. We do not
absorb the chemical potentials into the normalization factors since they
depend on time and describe the approach to thermal equilibrium.

The results of numerical simulations are reported below. We choose two
limiting initial conditions with flat spectra:\ a) electron-positron pairs
with a $10^{-5}$ energy fraction of photons and b) the reverse case, i.e.,
photons with a $10^{-5}$ energy fraction of pairs. Our grid consists of $60$
energy intervals and $16\times32$ intervals for two angles characterizing the
direction of the particle momenta. In both cases the total energy density is
$\rho=10^{24}$erg/cm$^{3}$. In the first case initial concentration of pairs
is $3.1\;10^{29}$cm$^{-3}$, in the second case the concentration of photons is
$7.2\;10^{29}$cm$^{-3}$.
\begin{figure}
[ptb]
\begin{center}
\includegraphics[
height=4.0811in,
width=3.4627in
]%
{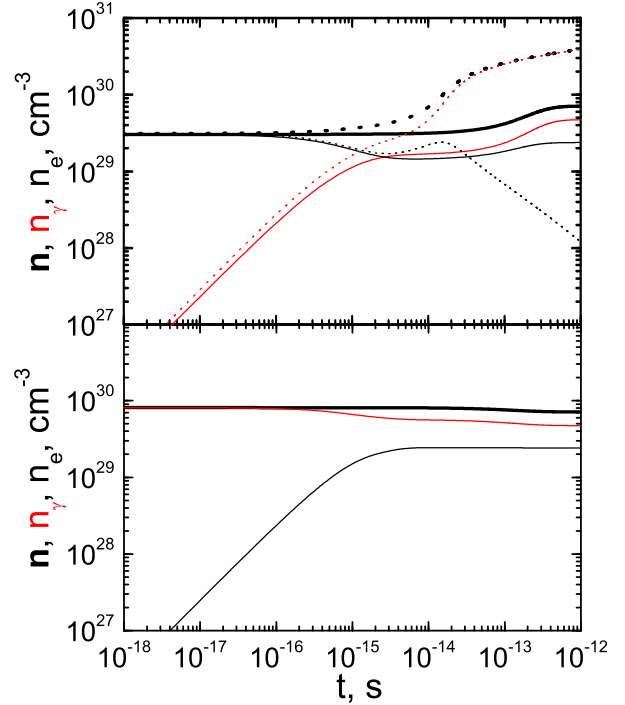}%
\caption{Depencence on time of concentrations of pairs (black), photons (red)
and both (thick) when all interactions take place (solid). Upper (lower)
figure corresponds to the case when initially there are mainly pairs
(photons). Dotted curves on the upper figure show concentrations when inverse
triple interactions are neglected. In this case an enhancement of the pairs
occurs with the corresponding increase in photon number and thermal
equilibrium is never reached.}%
\label{ntot}%
\end{center}
\end{figure}

In Fig.~\ref{ntot} we show concentrations of photons and pairs as well as
their sum for both our initial conditions. After calculations begin,
concentrations and energy density of photons (pairs) increase rapidly with
time, due to annihilation (creation) of pairs by the reaction {$\gamma
\gamma^{\prime}{\leftrightarrow}e^{\pm}e^{\mp}$. Then, in the kinetic
equilibrium phase, concentrations of each component stay almost constant, and
the sum of concentrations of photons and pairs remains unchanged. Finally,
both individual components and their sum reach stationary values. If one
compares and contrasts both cases as reproduced in Fig.~\ref{ntot} one can see
that, although the initial conditions are drastically different, in both cases
the same asymptotic values of the concentration are reached.}

\begin{figure}
[ptb]
\begin{center}
\includegraphics[
trim=0.000000in 0.007289in 0.000000in -0.007289in,
height=4.1537in,
width=3.3849in
]%
{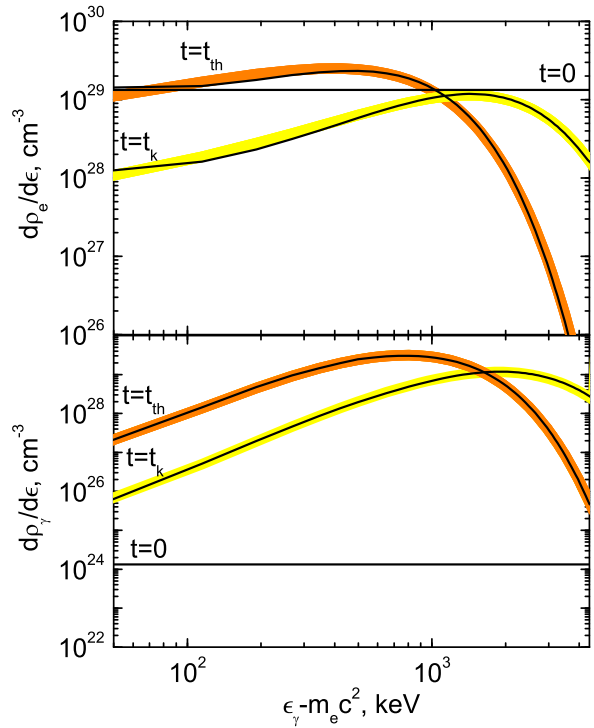}%
\caption{Spectra of pairs (upper figure) and photons (lower figure) when
initially only pairs are present. The black curve represents the results of
numerical calculations obtained successively at $t=0$, $t=t_{k}$ and
$t=t_{th}$ (see the text). Both spectra of photons and pairs are initially
taken to be flat. The yellow curves indicate the spectra obtained form
(\ref{dk}) at $t=t_{k}$. The perfect fit of the two curves is most evident in
the entire energy range leading to the first determination of the temperature
and chemical potential both for pairs and photons. The orange curves indicate
the final spectra as thermal equilibrium is reached.}%
\label{sp}%
\end{center}
\end{figure}

\begin{figure}
[ptb]
\begin{center}
\includegraphics[
height=4.0923in,
width=3.506in
]%
{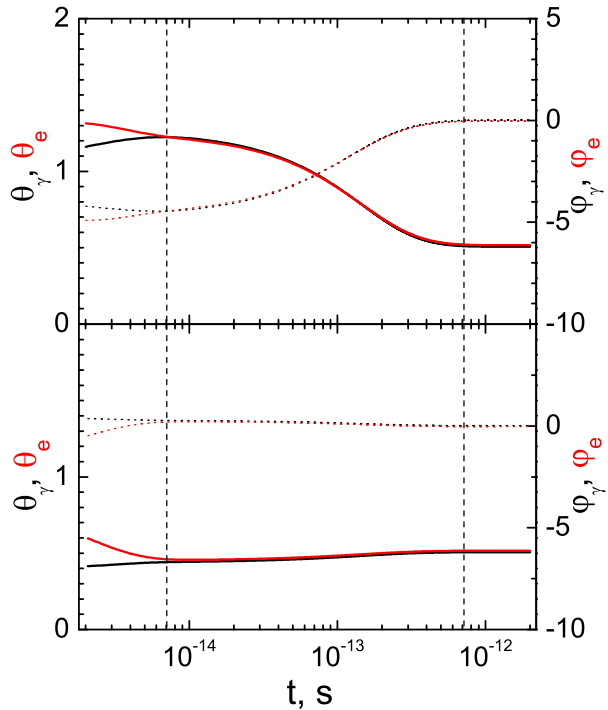}%
\caption{Time dependence of temperatures, measured on the left axis (solid),
and chemical potentials, measured on the right axis (dotted), of electrons
(black) and photons (red). The dashed lines correspond to the reaching of the
kinetic ($\sim10^{-14}$sec) and the thermal ($\sim10^{-12}$sec) equilibria.
Upper (lower) figure corresponds to the case when initially there are mainly
pairs (photons).}%
\label{thph}%
\end{center}
\end{figure}

We now describe in detail the case when initially pairs dominate. One can see
in Fig.~\ref{sp}\ that the spectral density of photons and pairs
\cite{Aksenov2004}
\begin{equation}
\frac{d\rho_{i}}{d\varepsilon}=\frac{4\pi}{c^{3}}f(\epsilon,t)\epsilon
^{3}\beta_{i},\label{spd}%
\end{equation}
where $\beta_{\pm}=\sqrt{1-(m_{e}c^{2}/\epsilon)^{2}}$ for pairs and
$\beta_{\gamma}=1$ for photons, can be fitted already at $t_{k}\approx
20t_{cs}\simeq7\;10^{-15}$sec by distribution functions (\ref{dk}) with
definite values of temperature $\theta_{k}(t_{cs})\approx1.2$ and chemical
potential $\varphi_{k}(t_{k})\approx-4.5$, common for pairs and photons. As
expected, after $t_{k}$ the distribution functions preserve their form
(\ref{dk}) with the values of temperature and chemical potential changing in
time, as shown in Fig.~\ref{thph}. As one can see from Fig.~\ref{thph} the
chemical potential evolves with time and reaches zero at the moment
$t_{th}\approx\alpha^{-1}t_{k}\simeq7\;10^{-13}$sec, corresponding to the
final stationary solution.%


We now discuss the results. Let us consider the distribution functions
(\ref{dk}) with different temperatures $\theta_{i}$\ and chemical potentials
$\varphi_{i}$ for pairs and photons. The requirement of vanishing reaction
rate for the Compton scattering $f_{\pm}f_{\gamma}=f_{\pm}{^{\prime}}%
f_{\gamma}{^{\prime}}$ leads to the equal temperature of pairs and photons
$\theta_{\pm}=\theta_{\gamma}\equiv\theta_{k}$, see also \cite{Pilla1997}%
,\cite{Ehlers1973}. In this way the detailed balance between any direct and
the corresponding inverse reactions shown in Tab.~\ref{tab1} leads to
relations between $\theta$ and $\varphi$ collected in Tab.~\ref{tab2}.

These relations are not imposed, but are verified through the numerical
calculations. This is a powerful tool to verify the consistency of our
approach and numerical calculations. These relations were obtained for the
first time, to our knowledge, in \cite{Ehlers1973} and then later in
\cite{Pilla1997} for binary reactions.%

\begin{table}[tbp] \centering
\begin{tabular}
[c]{|c|c|}\hline
Interaction & Parameters of distribution functions\\\hline\hline
Compton scattering & $\theta_{\gamma}=\theta_{\pm}$, $\forall\varphi_{\gamma}%
$,$\varphi_{\pm}$\\\hline
Pair production & $\varphi_{\gamma}=\varphi_{\pm}$, if $\theta_{\gamma}%
=\theta_{\pm}$\\\hline
Tripe interactions & $\varphi_{\gamma}$, $\varphi_{\pm}=0$, if $\theta
_{\gamma}=\theta_{\pm}$\\\hline
\end{tabular}
\caption{Relations between parameters of equilibrium DFs fulfilling detailed balance conditions for each of the reactions shown in Tab.~\ref{tab1}.}\label{tab2}
\end{table}%

From Tab.~\ref{tab2} one can see that the necessary condition for thermal
equilibrium in the pair plasma is detailed balance between direct and inverse
triple interactions. This point is usually neglected in the literature where
there are claims that the thermal equilibrium may be established with only
binary interactions \cite{Stepney1983}. In order to demonstrate it explicitly
we also show in Fig.~\ref{ntot} the dependence of concentrations of pairs and
photons when inverse triple interactions are artificially switched off. In
this case, see dotted curves in the upper Fig.~\ref{ntot}, after kinetic
equilibrium is reached concentrations of pairs decrease monotonically with
time, and thermal equilibrium is never reached.

The existence of a non-null chemical potential for photons indicates the
departure of the distribution function from the one corresponding to thermal
equilibrium. Negative (positive) value of the chemical potential generates an
increase (decrease) of the number of particles in order to approach the one
corresponding to the thermal equilibrium state. Then, since the total number
of particles increases (decreases), the energy is shared between more (less)
particles and the temperature decreases (increases), see fig. \ref{thph}.
Clearly, as in thermal equilibrium is approached, the chemical potential of
photons is zero.

In our example with the energy density $10^{24}$erg/cm$^{3}$ the thermal
equilibrium is reached at $\sim7\;10^{-13}$sec with the final temperature
$T_{th}=\allowbreak0.26$ MeV. For a larger energy density the duration of the
kinetic equilibrium phase, as well as of the thermalization timescale, is
smaller. In our entire temperature range (\ref{temp}) we deal with a
non-degenerate plasma.

Our results, obtained for the case of an uniform plasma, can only be adopted
for a description of a physical system with dimensions $R_{0}\gg\frac
{1}{n\sigma_{T}}=4.3\;10^{-5}$cm.

The assumption of the constancy of the energy density is only valid if the
dynamical timescale $t_{dyn}=\left(  \frac{1}{R}\frac{dR}{dt}\right)  ^{-1}$
of the plasma is much larger than the above timescale $t_{th}$ which is indeed
true in all the cases of astrophysical interest.

Since we get thermal equilibrium already on the timescale $t_{th}%
\lesssim10^{-12}$sec, and such a state is independent of the initial
distribution functions for electrons, positrons and photons, the sufficient
condition to obtain an isothermal distribution on a causally disconnected
spatial scale $R>ct_{th}=10^{-2}$cm is the request of constancy of the energy
density on such a scale as well as, of course, the invariance of the physical laws.

We have considered the evolution of an initially nonequilibrium optically
thick electron-positron-photon plasma up to reaching thermal equilibrium.
Starting from arbitrary initial conditions we obtain kinetic equilibrium from
first principles, directly solving the relativistic Boltzmann equation with
collisional integrals computed from QED matrix elements. We have demonstrated
the essential role of direct and inverse triple interactions in reaching
thermal equilibrium.\ Our results can be applied in the theories of the early
universe and of gamma-ray bursts, where thermal equilibrium is postulated at
the very early stages. These results can in principle be tested in laboratory
experiments in the generation of electron-positron pairs.

\begin{acknowledgments}
We thank the anonymous referee for comments which have improved the
comprehension of our results.
\end{acknowledgments}


\begin{thebibliography}{99}                                                                                               %


\bibitem {Kolb1990}E.W. Kolb and M.S. Turner, The Early Universe, Perseus
Books Group (1993).

\bibitem {Piran1999}J. Goodman, ApJ 308, L47 (1986); T. Piran,
Phys.\ Rep.\ 314, 575 (1999); R. Ruffini et al., A\&A 350, 334 (1999); A\&A
359, 855 (2000).

\bibitem {Wardle1998}J.F.C. Wardle et al., Nature 395, 457 (1998).

\bibitem {Churazov2005}E. Churazov et al., MNRAS 357, 1377 (2005).

\bibitem {Usov1998}V.V. Usov, Phys.\ Rev.\ Lett., 80, 230 (1998).

\bibitem {Blaschke2006}D.B. Blaschke et al., Phys.\ Rev.\ Lett.\ 96, 140402 (2006).

\bibitem {Bisnovatyi1971}G.S. Bisnovatyi-Kogan, Y.B. Zel'dovich, and R.A.
Syunyaev, Soviet Astronomy 15, 17 (1971); A.P. Lightman, ApJ 253, 842 (1982);
R. Svensson, ApJ 258, 335 (1982); P.W. Guilbert and S. Stepney, MNRAS 212, 523
(1985); P.S. Coppi, R.D. Blandford, MNRAS 245, 453 (1990); S. Iwamoto, F.
Takahara, ApJ 601, 78 (2004).

\bibitem {BB1956}S.T. Belyaev and G.I. Budker, DAN SSSR 107, 807 (1956)
[Sov.\ Phys.\ Dokl.\ 1, 218 (1956)]; D. Mihalas, Foundations of Radiation
Hydrodynamics, Oxford (1984).

\bibitem {Aksenov2004}A.G. Aksenov, M. Milgrom, and V.V. Usov, ApJ 609, 363 (2004).

\bibitem {Hall1976}G. Hall, J.M. Watt, Modern Numerical Methods for Ordinary
Differential Equations, Oxford (1976).

\bibitem {Berestetskii1982}E.M. Lifshitz, L.P. Pitaevskii, and V.B.
Berestetskii, Quantum Electrodynamics, Elsevier (1982); W. Greiner, J.
Reinhardt, Quantum Electrodynamics, Springer (2003); A.I. Akhiezer, V.B.
Berestetskii, Quantum Electrodynamics, Nauka (1981).

\bibitem {Svensson1984}R. Svensson, MNRAS 209, 175 (1984).

\bibitem {Pilla1997}R. P. Pilla and J. Shaham, ApJ 486, 903 (1997).

\bibitem {Ehlers1973}J. Ehlers, in Relativity, Astrophysics and Cosmology, ed.
W. Israel, Reidel (1973).

\bibitem {Stepney1983}S. Stepney, MNRAS 202, 467 (1983).
\end{thebibliography}
\end{document}